# Raman evidence for coupling of superconducting quasi-particles with a phonon and crystal field excitation in superconductor $Ce_{0.6}Y_{0.4}FeAsO_{0.8}F_{0.2}$

Pradeep Kumar[1], D. V. S. Muthu[1], J. Prakash[2], A. K. Ganguli[2] and A. K. Sood[1,*]

[1]Department of Physics, Indian Institute of Science, Bangalore -560012, India

[2]Department of Chemistry, Indian Institute of Technology, New Delhi -110016, India

**ABSTRACT**

We report inelastic light scattering experiments on superconductor $Ce_{0.6}Y_{0.4}FeAsO_{0.8}F_{0.2}$ from 4K to 300K covering the superconducting transition temperature $T_c \sim 48.6K$. A strong evidence of the superconductivity induced phonon renormalization for the $A_{1g}$ phonon mode near 150 cm$^{-1}$ associated with the Ce/Y vibrations is observed as reflected in the anomalous red-shift and decrease in the linewidth below $T_c$. Invoking the coupling of this mode with the superconducting gap, the superconducting gap ($2\Delta$) at zero temperature is estimated to be ~ 20 meV i.e the ratio $2\Delta(0)/\kappa_B T_c$ is ~ 5, suggesting $Ce_{0.6}Y_{0.4}FeAsO_{0.8}F_{0.2}$ to belong to the class of strong coupling superconductors. In addition, the mode near 430 cm$^{-1}$ associated with $Ce^{3+}$ crystal field excitation also shows anomalous increase in its linewidth below $T_c$ suggesting strong coupling between crystal field excitation and the superconducting quasi-particles. Our observations of two high frequency modes (S9 and S10) evidence the non-degenerate nature of $Fe^{2+}$ $d_{xz/yz}$ orbitals suggesting the electronic nematicity in these systems.

*Corresponding author: email: asood@physics.iisc.ernet.in, Ph: +91-80-22932964

# 1. INTRODUCTION

The discovery of superconductivity in RFeAsO$_{1-x}$F$_x$ (R = La, Sm, Ce, Nd, Pr and Gd ) [1-3] with a superconducting transition temperature ($T_c$) up to 55 K has generated enormous interest to investigate these materials both experimentally and theoretically. Especially debated is the question whether or not these materials share common pairing mechanism with the cuprates and conventional low temperatures superconductors. There are already a numbers of different scenarios proposed for the possible pairing mechanism in these materials [4-9] but the microscopic mechanism of pairing is still elusive. One of the most important quantity for understanding the pairing mechanism in high temperature superconductors is the superconducting gap ($2\Delta$), whose magnitude and structure are linked to the pairing mechanism.

In superconductors, the opening of the superconducting gap below $T_c$ renormalizes the electronic states near the Fermi surface which, in turn, can change the self energies of some of the phonons i.e their frequencies (real part of the self energy) and linewidths (imaginary part of the self energy) can be affected drastically. In case of cuprates and iron-based superconductors (FeBS), Raman and Infrared spectroscopy have revealed anomalies in phonon parameters below $T_c$ attributed to the coupling of superconducting quasi-particles excitations with the phonons [10-16]. The changes in phonon self-energies have been used to probe the magnitude and even the symmetry of the superconducting order parameter [10, 17]. In addition, Raman spectroscopy has also been used to probe the coupling of the phonons mode with the crystal field excitations in high temperature superconductors. Raman studies on iron-based superconductors cover "42622" [13], "1111" [14, 18-22], "122" [15, 23-24] and "11" systems [25-26]. The phonon anomalies eluded above due to coupling with the superconducting quasi-particles have been observed in case of $CeFeAsO_{0.9}F_{0.1}$ [14], the $E_g$ (389 $cm^{-1}$) phonon mode associated with the oxygen vibrations shows coupling with $Ce^{3+}$ crystal field excitations.

Recently, it has been shown that doping of *Y* in $CeFeAsO_{1-x}F_x$ increases its superconducting transition temperature from 42.7 K to 48.6 K [27]. There is no Raman study of *Y*-doped $Ce_{1-\delta}Y_\delta FeAsO_{1-x}F_x$ system either at room-temperature or as a function of temperature. In this paper we report such a study of $Ce_{0.6}Y_{0.4}FeAsO_{0.8}F_{0.2}$ with $T_c \sim$ 48.6 K, focusing mainly on the temperature dependence of the lowest frequency phonon mode and coupling of crystal field excitation with the superconducting quasi-particles below $T_c$.

## 2. RESULTS AND DISCUSSIONS

### 2.1. Raman Scattering from Phonons

CeFeAsO has a layered structure belonging to the tetragonal *P4/nmm* space group [21]. There are eight Raman active phonon modes belonging to the irreducible representation $2A_{1g} + 2B_{1g} + 4E_g$ [21]. Figure 1 shows Raman spectrum at 4 K, revealing 10 modes labeled as S1 to S10 in the spectral range 100 - 1800 $cm^{-1}$. Spectra are fitted to a sum of Lorentzian functions. The individual modes are shown by thin lines and resultant fit by thick lines. The frequencies measured at 4 K are tabulated in Table I. The existing experiments and theoretical calculation [14, 18-26, 28] of phonons show that first order Raman phonons are observed only below 550 cm$^{-1}$. However we clearly observed three modes (S8, S9 and S10) above 600 cm$^{-1}$. Following the earlier Raman studies on "1111" systems [14, 18, 20-22], we assign the modes S1 to S5 to be the first-order Raman active modes as S1 (~ 146 cm$^{-1}$, $A_{1g}$, Ce/Y), S2 ( ~ 220 cm$^{-1}$, $B_{1g}$, Fe), S3 ( ~ 255 cm$^{-1}$, $E_g$, Fe), S4 ( ~ 290 cm$^{-1}$, $B_{1g}$, O) and S5 ( ~ 394 cm$^{-1}$, $E_g$, O) (see Table I ). The mode S6 is assigned to the crystal field excitation of $Ce^{3+}$ (to be explained later), S7 and S8 as second-order Raman modes and S9, S10 as crystal field excitations of $Fe^{2+}$ d-levels.

**2.2. Temperature Dependence of the Phonon Frequencies**

The anomalous change in the frequency and linewidth of the phonons across the superconducting transition temperature can be attributed to change in the self energies, $\Delta\Sigma = \Delta\omega - i\Delta\Gamma$, induced by the superconducting transition [29]. The real and imaginary parts of the self energy renormalize frequency and linewidth, respectively. Within the framework of strong coupling Eliasberg theory, Zeyer et al [29] showed that a change in the phonon self energy across transition temperature is linked with the interaction of the phonons with the superconducting quasiparticles excitations. Qualitatively, phonons below the superconducting gap ($2\Delta$) become sharp and show frequency softening whereas the phonons above the gap may broaden and show anomalous hardening. The renormalization of a given phonon mode is strongest when the phonon energy is close to the gap. Figure 2(a) shows the frequency and linewidth of mode S1 as a function of temperature. Figure 2(b) shows a few typical spectra in the range of S1 mode to reflect the anomalous behaviour shown in fig. 2(a). The anomalous sharpening and frequency softening of the mode below $T_c$ indicates a strong coupling of this phonon of $A_{1g}$ symmetry involving the vibration of Ce/Y with the electronic states. A microscopic understanding of this coupling can help in understanding the symmetry of the superconducting gap. Equaling the phonon energy with the gap energy at zero degree, this gives an estimate of $2\Delta(0)/\kappa_B T_c \sim 5$, suggesting this system to be a strong-coupling superconductor. We note that in "1111" systems experimental evidence of single and multiple gaps have already been reported [14, 30-33], where ratio varies from 3 to 10. The origin of this large variation is still not understood.

In comparison to mode S1, Fig. 3(a) shows that temperature dependence of the modes S5, S7 and S8 is normal as expected based on anharmonic and quasi-harmonic effects. The solid line for the first-order mode S5 is fitted to an expression based on a simple model of cubic anharmonicity where the phonon mode decays into two equal energy phonons [34]: $\omega(T) = \omega(0) + C[1 + 2n(\omega(0)/2)]$, where $\omega(T)$ and $\omega(0)$ are the phonon frequencies at temperature T and T = 0 K, C is Self-energy parameter for a given phonon mode and $n(\omega) = 1/[\exp(\hbar\omega/\kappa_B T) - 1]$ is the Bose-Einstein mean occupation factor. Fitted parameters for the mode S5 are $\omega(0)$ = 396.9 cm$^{-1}$, C = -5.4 ± 1.4 cm$^{-1}$.

**2.3. Raman Scattering from Crystal Field Split Excitations of Ce$^{3+}$ and Origin of the High Frequency Modes S8, S9 and S10**

In earlier temperature-dependent Raman study of superconductor CeFeAsO$_{0.9}$F$_{0.1}$ [14] a mode near 430 cm$^{-1}$ was observed and was assigned to the electronic Raman scattering involving the crystal field excitations (CFE) of Ce$^{3+}$ f-levels (J = 5/2). The frequency of mode S6 (not shown here) remains nearly constant with temperature similar to that in earlier studies. Following earlier studies we have assigned mode S6 to the CFE of Ce$^{3+}$.

Fig. 3(b) shows the temperature dependence of the full width at half maxima (FWHM) of the phonon mode S5 and the CF excitation S6. It can be seen that the linewidth of mode S5 is as expected from anharmonic effects, fitted by a solid line $\Gamma(T) = \Gamma(0) + A[1 + 2n(\omega(0)/2)]$, [34] where A is a constant and $\Gamma(0)$ is intrinsic linewidth at zero Kelvin. In comparison, the linewidth of the mode S6 is anomalous below $T_c$, reflecting a strong coupling of the CF excitation of $Ce^{3+}$ with the superconducting quasi-particles. This is also corroborated by the increase in the normalized intensity of the mode S6 below $T_c$ (see fig. 3(c)). The coupled CFE with the longitudinal optical phonons have been observed in case of $CeFeAsO_{0.9}F_{0.1}$ and $UO_2$ [14, 35]. Following this, the mode S8 (860 cm$^{-1}$) is assigned as a second-order Raman scattering from a combination of modes S5 and S6.

We also observed two high energy excitations at 1354 cm$^{-1}$ (S9) and 1592 cm$^{-1}$ (S10). Our recent Raman studies on $CeFeAsO_{0.9}F_{0.1}$ and $FeSe_{0.82}$ superconductors have also shown similar modes and were attributed to electronic Raman scattering between crystal field split *d*-orbitals of $Fe^{2+}$ as ($x^2$-$y^2$) level to *xz* and *yz* levels [14, 25]. We make this assignment in the present case as well, giving us a value of splitting of *d*-orbitals *xz* and *yz* as ~ 29 meV. We note that the recent experimental as well theoretical studies on FeBS suggest that orbital and spin degrees of freedom are believed to conspire to give much elusive glue for the cooper pair formation [36-37]. In particular, Fe $t_{2g}$ ($d_{xy}$, $d_{xz/yz}$) orbitals play a central role in controlling their electronic as well as magnetic properties. Further, reports on FeBS have shown the existence of electronic nematicity between Fe $d_{xz}$ and $d_{yz}$ orbitals [14, 25, 38] evidencing the crucial role of orbital degrees of freedom.

## 3. CONCLUSIONS

In conclusion, the anomalous behavior of the Raman active phonon mode (S1) below $T_c$ is attributed to the strong coupling between the phonon mode and superconducting quasi-particles excitations due to the opening of superconducting gap below $T_c$. The Raman mode near 430 cm$^{-1}$ associated with the CFE of Ce$^{3+}$ is also shown to be coupled strongly with the superconducting quasi-particle excitations. Our results obtained here suggest that the strong interplay phonon and electronic degree of freedom is important to understand the pairing mechanism in iron-based superconductors.

## Acknowledgments

PK and JP acknowledge CSIR, India, for research fellowship. AKS and AKG acknowledge the DST, India, for financial support.

Table-I: List of the experimental observed frequencies at 4K in Ce$_{0.6}$Y$_{0.4}$FeAsO$_{0.8}$F$_{0.2}$.

| Mode Assignment | Experimental $\omega$ (cm$^{-1}$) |
|---|---|
| S1  A$_{1g}$ (Ce) | 146 |
| S2  B$_{1g}$ (Fe) | 220 |
| S3  E$_g$ (Fe) | 255 |
| S4  B$_{1g}$ (O) | 290 |
| S5  E$_g$ (O) | 394 |
| S6  CF of Ce$^{3+}$ levels | 430 |
| S7  Two-Phonon | 534 |
| S8  (S5 + S6) | 860 |
| S9  CF of Fe-d levels | 1354 |
| S10 CF of Fe-d levels | 1592 |

**FIGURE CAPTION**

FIG.1. (Color online) Unpolarised Raman spectra of $Ce_{0.6}Y_{0.4}FeAsO_{0.8}F_{0.2}$ at 4 K. Thin solid lines are Lorentzian fit to individual modes and the solid thick line shows the total fit to the experimental data.

FIG.2. (Color online) (a) Temperature-dependence of phonon mode S1. (b) Temperature evolution of mode S1 at a few temperatures.

FIG.3. (Color online) (a) Temperature dependence of the modes S5, S7 and S8. The solid line for mode S5 is the fitted curve as described in text. (b) Normalized integrated intensity of mode S6. Solid line is linear fit in two temperature ranges. (c) Temperature dependence of the linewidth of modes S5 and S6. Solid line for mode S6 is linear fit in two temperature ranges and for mode S5 is fitted curve as described in text.

FIGURE 1:

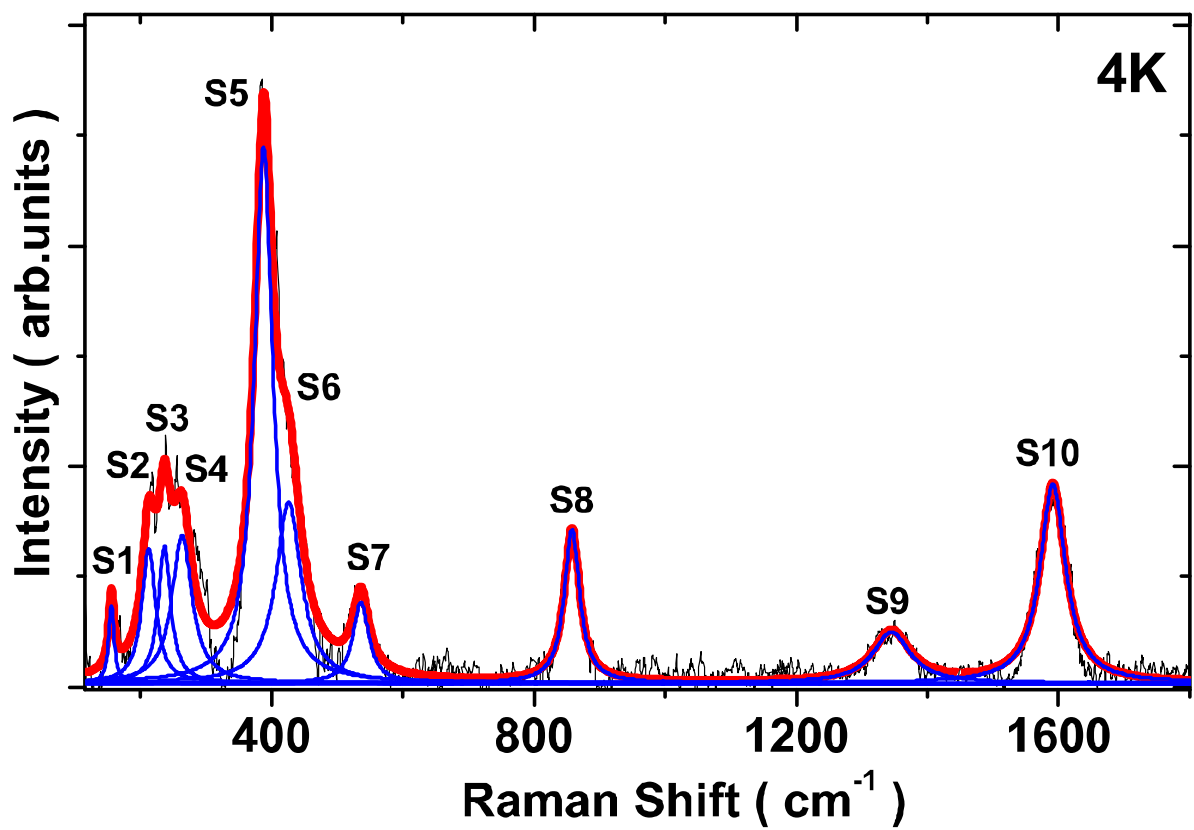

FIGURE 2:

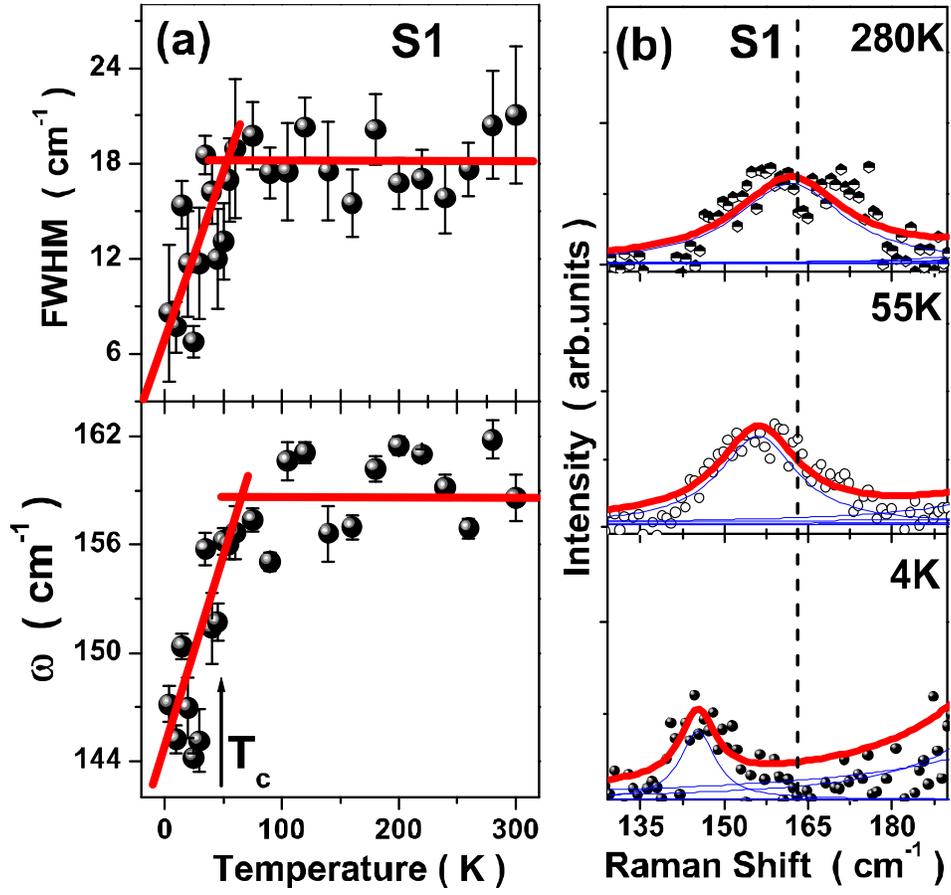

FIGURE 3 :

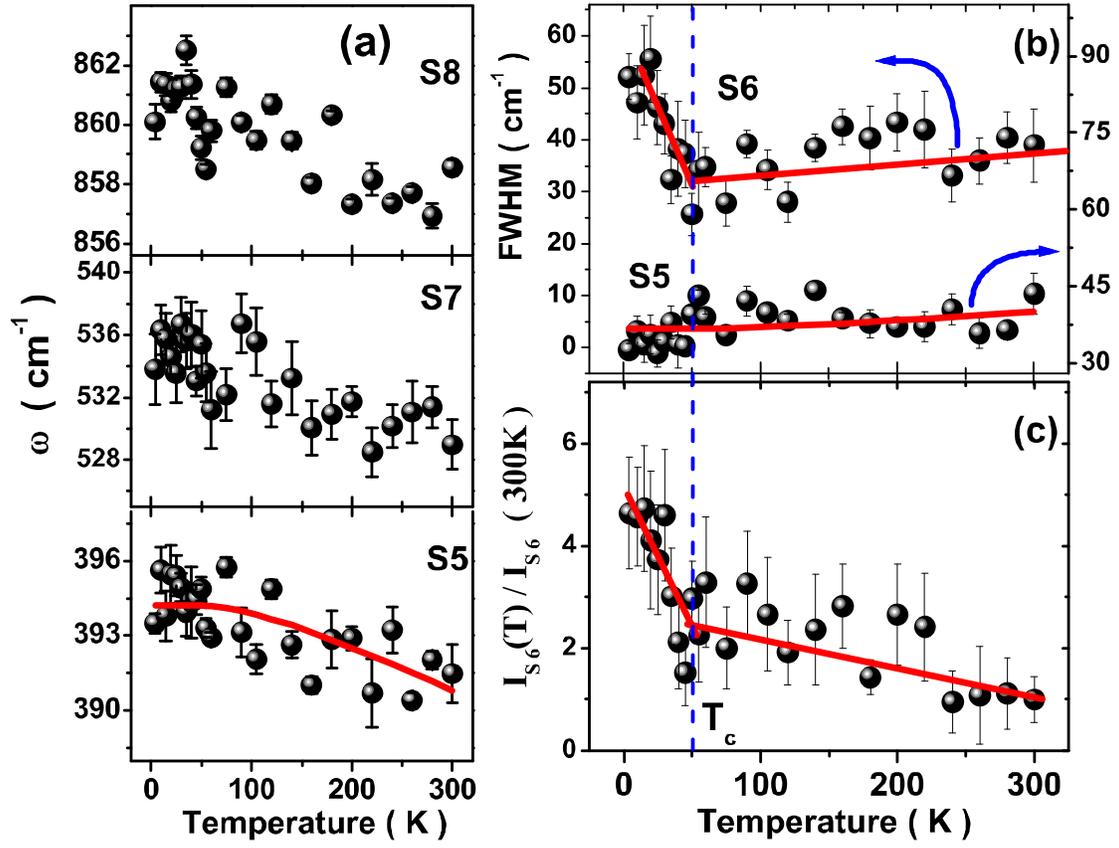